# Multipiezo effect in altermagnetic V$_2$SeTeO monolayer

Yu Zhu,[1,ξ] Taikang Chen,[1,ξ] Yongchang Li,[1] Lei Qiao,[1] Xiaonan Ma,[1] Tao Hu,[2] Heng Gao,[1,*] and Wei Ren[1,3,‡]

[1]*Department of Physics, Shanghai Key Laboratory of High Temperature Superconductors, International Centre of Quantum and Molecular Structures, Shanghai University, Shanghai 200444, China*
[2]*School of Materials Science and Engineering, Shanghai University, Shanghai 200444, China*
[3] *Zhejiang Laboratory, Hangzhou 311100, China*

[*]gaoheng@shu.edu.cn  [‡]renwei@shu.edu.cn

[ξ]These authors contributed equally to this work.

Inspired by recent theoretical proposal on the interesting piezomagnetism and C-paired valley polarization in V$_2$Se$_2$O monolayer, we predict a stable antiferromagnetic Janus monolayer V$_2$SeTeO with altermagnetic configuration using density functional theory calculations. It exhibits a novel "multi-piezo" effect combining piezoelectric, piezovalley and piezomagnetism. Most interestingly, the valley polarization and the net magnetization under strain in V$_2$SeTeO exceed these in V$_2$Se$_2$O, along with the additional large piezoelectric coefficient of $e_{31}$ (0.322*10$^{-10}$ C/m). The "multi-piezo" effect makes antiferromagnetic Janus monolayer V$_2$SeTeO a tantalizing material for potential applications in nanoelectronics, optoelectronics, spintronics and valleytronics.

# I. INTRODUCTION

The pursuit of miniaturization for multifunctional electronic devices necessitates exploring new degrees of freedom and approaches to integrate and manipulate multiple functions in nanoscale materials [1-5]. The exfoliation of graphene in 2004 [6] sparked tremendous interest in two-dimensional (2D) materials such as boronene [7,8], phosphene [9,10], and monochalcogenides [11] , which can be exfoliated into monolayers possessing unique properties for quantum devices. Multiferroics, materials exhibiting coupled two or more ferroic orders (i.e., ferroelectric, ferromagnetic, and ferroelastic), are particularly interesting for next-generation of electronics [12-16]. However, stabilizing multiple ferroic orders in a 2D material remains challenging [17]. Thus, whether there are other alternatives to achieve the purpose of multifunctional nanomaterials?

Strain engineering provides a potential method for tuning and enriching physical properties thanks to the high stretchability of nanomaterials [18-20]. Under mechanical strain, nanomaterials exhibit interesting responses including piezoelectricity, piezoluminescence and piezomagnetism. Piezoelectricity is a well-known phenomenon of electromechanical coupling that can generate voltage under mechanical strain [21,22]. Piezoluminescence refers to light emission during elastic deformation [23,24]. Piezomagnetism occurs in some antiferromagnetic (AFM) crystals, where applied external stress induces ferromagnetic moment [25-27]. In contrast to conventional ferrovalley materials requiring time-reversal symmetry, recent works showed valleys in altermagnetic $V_2Se_2O$ [28] monolayer are found to be protected by crystal symmetry and coined C-paired spin-valley-locking (SVL). Applying uniaxial strain generates valley polarization, known as piezovalley effect. Additionally, Doping $V_2Se_2O$ can also induce piezomagnetism [28]. An intriguing question is whether piezoelectricity can be introduced into $V_2Se_2O$ while retaining the multi-piezo effect? Recent interest in 2D Janus structures [29-32] provides an opportunity, since they exhibit intrinsic mirror symmetry breaking that enables out-of-plane piezoelectric.

In our work, motivated by recent experimental progress in the synthesis of Janus monolayers [33,34] (MoSSe, for example) and bulk $V_2Se_2O$ [35], along with theoretical studies of $V_2Se_2O$ monolayer [28], we designed a Janus $V_2SeTeO$ monolayer. It naturally exhibits broken out-of-plane symmetry and intact in-plane symmetry, enabling large out-of-plane piezoelectricity while preserving the original in-plane properties. A novel multi-piezo effect combing piezovalley polarization, piezomagnetism and piezoelectricity is realized in the Janus $V_2SeTeO$ monolayer. Remarkably, these three mechanical responses are not causally related to each other, not as the phenomenon of magneto-electric coupling in multiferroics does. The modulation methods and mechanisms of these three piezo-effects are independent of each other, with piezoelectricity being out-of-plane, piezovalley and piezomagnetism being protected by in-plane symmetry. A ferromagnetic moment only arises upon doping the strained $V_2SeTeO$. Our discovery of the multi-piezo effect provides a new direction and theoretical guidance for designing multifunctional nanodevices which couple electrical, chemical and magnetic controls through external mechanical strain.

## II. COMPUTATIONAL METHODS

We performed calculations on the basis of density functional theory (DFT) by employing the Vienna ab initio simulation package (VASP) [36,37], and the projected augmented wave (PAW) method [38]. The cutoff energy for plane-wave expansion was set to be 600 eV. Regarding the exchange-correlation functional, the generalized gradient approximation (GGA) of the Perdew-Burke-Ernzerh was adopted [39]. The optB88b-vdW method was employed for van der Waals (vdW) correction during the structure optimization [40,41]. V-3d electron localization [42] was treated with GGA+U approach [43] using U = 4 eV, consistent with previous reports [44-46]. The crystal structure was fully relaxed with the convergence criteria for the total energy and force, which were set to be $10^{-8}$ eV and 0.005 eV/Å, respectively. The Monkhorst−Pack k-points were sampled in the 2D Brillouin zone using a Γ-centered 10 × 10 × 1 k-grid. To avoid interlayer interactions, a vacuum layer of 25 Å is placed along the c-axis. The ab initio molecular dynamics (AIMD) simulations [47] were carried out using 5 × 5 × 1 supercell at 300 K with a 1 fs time step. The phonon dispersion spectrum using density-functional perturbation theory (DFPT) with a 4×4×1 supercell was calculated to verify the dynamic stability using the PHONOPY code [48,49].

## III. RESULTS AND DISCUSSIONS

### A. Geometric and altermagnetism

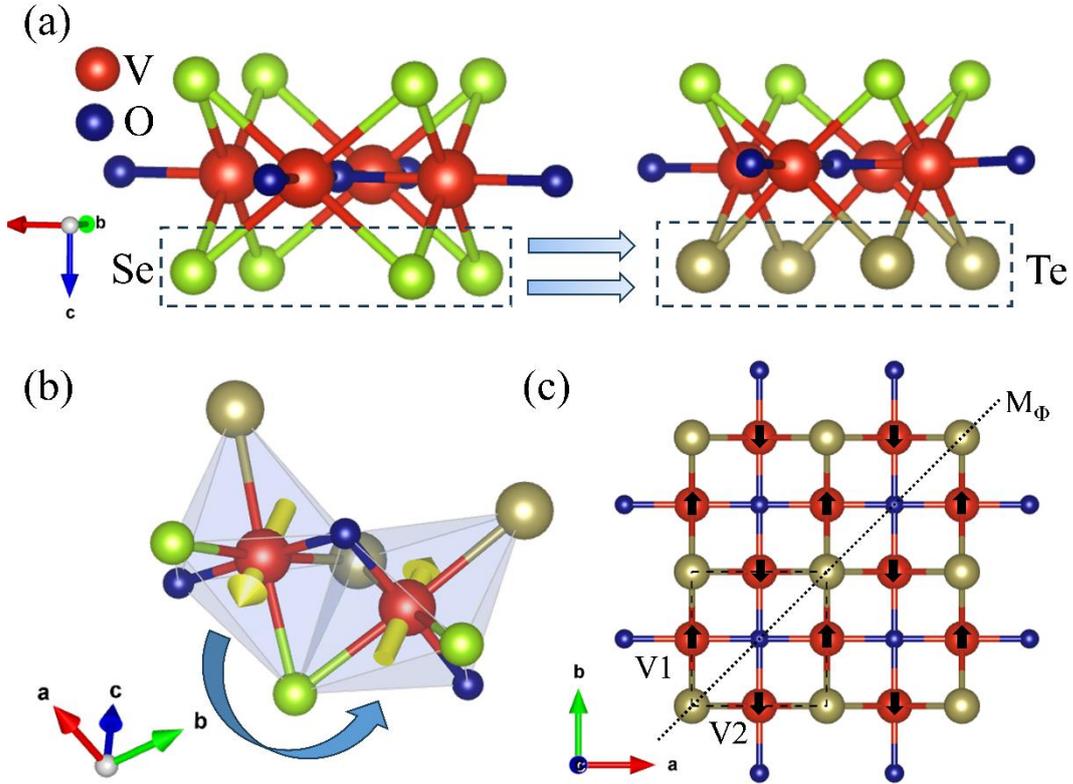

Figure 1 (a) Schematic diagram of the element substitution from $V_2Se_2O$ monolayer to Janus $V_2SeTeO$ monolayer. (b) Altermagnetic structure of $V_2SeTeO$ monolayer, where the yellow arrows directions indicate the spin alignment on the sublattice of the two V atoms. (c) The 2D Neel AFM order has magnetic moments located on the V1 and V2 atoms (black arrows). The dotted line shows the primitive cell. $M_\Phi$ denotes diagonal mirror symmetry in the ab plane.

$V_2Se_2O$ monolayer is a three-layer atomic structure sandwiched by two Se planes [28]. Substituting Se with Te in the bottom layer of $V_2Se_2O$ creates the Janus $V_2SeTeO$ monolayer, as depicted in Figure 1 (a). Compared to $V_2Se_2O$ monolayer (space group number 123, P4/mmm), Janus $V_2SeTeO$ monolayer (space group number 99, P4mm) exhibits lower symmetry due to out-of-plane mirror symmetry breaking while retaining the tetragonal structure and in-plane $C_{4v}$ symmetry. The calculated structural parameters including lattice constants, bond lengths and bond angles for both $V_2Se_2O$ and $V_2SeTeO$ monolayers are listed in Table S1. The length of the V-Se bond is shorter than that of the V-Te bond and the angles between Se-V-Se and Te-V-Te are different, while the anisotropic bonding shears the structure and reduces its symmetry. The $V_2SeTeO$ primitive cell contains two V atoms, and each of them is coordinated by two Se atoms, two Te atoms, and two oxygen atoms in a highly distorted octahedral structure in Figure 1 (b), possessing ~2 $\mu_B$ local magnetic moments. This indicates that the Janus structure primarily impacts on the symmetry without significantly changing the magnetic properties. The in-plane diagonal mirror symmetry $M_\Phi$ is preserved, changing from the Se-O-Se to Se/Te-O-Se/Te plane as shown in Figure 1 (c).

Four magnetic configurations of a 2×2×1 Janus $V_2SeTeO$ supercell (Figure S1) were constructed and calculated to determine the magnetic ground state. AFM-Neel configuration (Figure 1 (c)) is the most stable, with energy differences listed in Table I, confirming the intrinsic antiferromagnetism akin to $V_2Se_2O$. Interestingly, this AFM Neel state exhibits the recently defined altermagnetism [50,51], enabling spin splitting in conventional collinear antiferromagnets. To better characterize the magnetic properties, the calculated magnetic anisotropy energy (MAE) is found to be 143 μeV/unit cell, with the magnetic easy-axis along the diagonal of the *ab* plane.

Figure S2 (a) presents the calculated phonon spectrum for the $V_2SeTeO$ monolayer and there is no obvious imaginary frequency, which confirms the structural stability. The thermal stability of Janus $V_2SeTeO$ was also verified through *ab initio* molecular dynamics with only slight energy fluctuations during the simulations (Figure S2 (b)). The elastic constants (Table S2) satisfy the Born criteria [52]: $C_{11} > 0$, $C_{66} > 0$, and $C_{11} > |C_{12}|$ indicating the mechanical stability of $V_2SeTeO$ monolayer. In summary, these results demonstrate the intrinsic stability of the Janus $V_2SeTeO$ monolayer in the AFM ground state.

**B. Electronic properties and piezovalley**

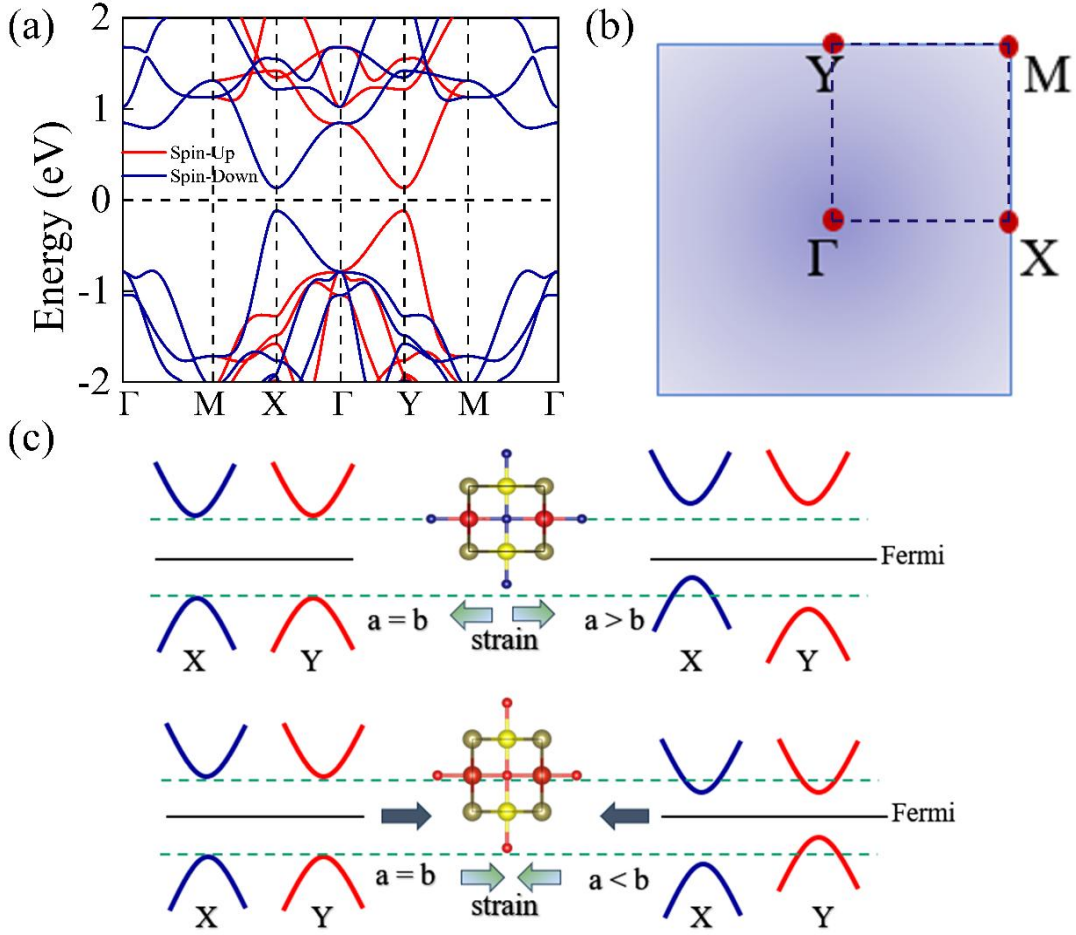

Figure 2 (a) Spin-resolved electronic band structure of Janus $V_2$SeTeO monolayer, with spin-up (red lines) and spin-down (blue lines) channels. (b) The first Brillouin zone with high symmetry points. (c) Schematic of the piezovalley mechanism, the blue and red lines indicate the valleys at points X and Y, respectively.

The band structure and density of states (DOS) of $V_2$SeTeO monolayer are shown in Figure 2 (a) and Figure S3 by GGA+U method without spin-orbital coupling (SOC). The Brillouin zone for band structure calculations is shown in Figure 2 (b), which is related to the crystal structure. It turns out that the $V_2$SeTeO monolayer is a semiconductor with a 0.25 eV direct band gap, smaller than the value 0.72 eV of $V_2Se_2O$ [28], owing to the more metallic nature of Te versus Se. Notably, the band structure in Figure 2 (a) exhibits clear spin-splitting, distinct from conventional AFM materials. This arises from its altermagnetic order shown in Figure 1 (b)-(c), which breaks both PT symmetry (the combination of spatial inversion P and time reversal T symmetries) and UL symmetry (the combination of spinor U and translation L symmetries) to allow spin splitting with or without SOC [53]. (See Note S1 for the detailed symmetry rules). Altermagnetism has also been reported in many known materials, like bulk oxide $MnO_2$ [54] and $RuO_2$ [55,56], perovskite oxide insulators $LaMnO_3$ [57,58] and $CaCrO_3$ [59], 2D materials $V_2Se_2O$ [28] and CrO [60]. As shown in Figure 1 (b)-(c), in the alternating magnetic configuration, the magnetic V atoms with different spin directions cannot be related by pure translation due to differing spin directions, requiring instead rotation operations that break symmetry and induce spin polarization. Exploiting this feature of altermagnetism may overcome the inability for spin polarization in AFMs, better enabling spintronic applications.

Previous works on $V_2Se_2O$ [28] showed uniaxial strain can produce valley polarization in the altermagnetic structure. Similarly, degenerate band valleys are observed at X and Y points in $V_2SeTeO$ (Figure 2 (a)), contributed by V atoms with different spin alignments. This exemplifies the SVL, lock-coupling the spin and valley degrees of freedom. Valley traditionally arises in transition metal dichalcogenides (TMDs) via ferrovalley ordering, requiring strong SOC, broken inversion symmetry and heavy elements [61-63]. Uniquely here, valleys in $V_2SeTeO$ are protected by the mirror symmetry rather than the time-reversal symmetry. This relaxes the constraints for realizing valley polarization, providing new avenues to design valleytronics materials.

As in $V_2Se_2O$, in-plane uniaxial strain along the a/b axis breaks the mirror symmetry and generates a valley polarization which we term the "piezovalley effect". The physical mechanism of piezovalley can be seen in Figure 2 (c). The valley energies E(X) and E(Y) at X and Y are no longer degenerate, thus the valley polarization can be defined as the energy difference between two valleys P=E(X)-E(Y). The valley polarization of valence and conduction bands versus uniaxial strain along *a* direction is shown in Figure 3 (a). It turns out that the valley polarization changes monotonously under the control of strain, with opposite trends along the *a* and *b* directions due to the diagonal mirror symmetry $M_\Phi$. Remarkably, the valley polarization of 128 meV induced by 4% tensile strain exceeds reported values in 2D ferrovalley materials such as LaBrI (59 meV) [64], 2H-$FeCl_2$ (101 meV) [65], VSCl (57.8 meV) [66], and VSSe (85 meV) [67], and also larger than ~ 60 meV in $V_2Se_2O$ [28]. As for electronic properties, the band structures of $V_2SeTeO$ under uniaxial strain from -4% to 4% are plotted in Figure S4, as well as band gap variation (two valleys' band gaps [X and Y] and overall gap [total]) in Figure 3 (b). These results indicate that the predicted Janus $V_2SeTeO$ monolayer retains altermagnetic structure and realize notably enhanced piezovalley under stain.

**C．Piezomagnetism**

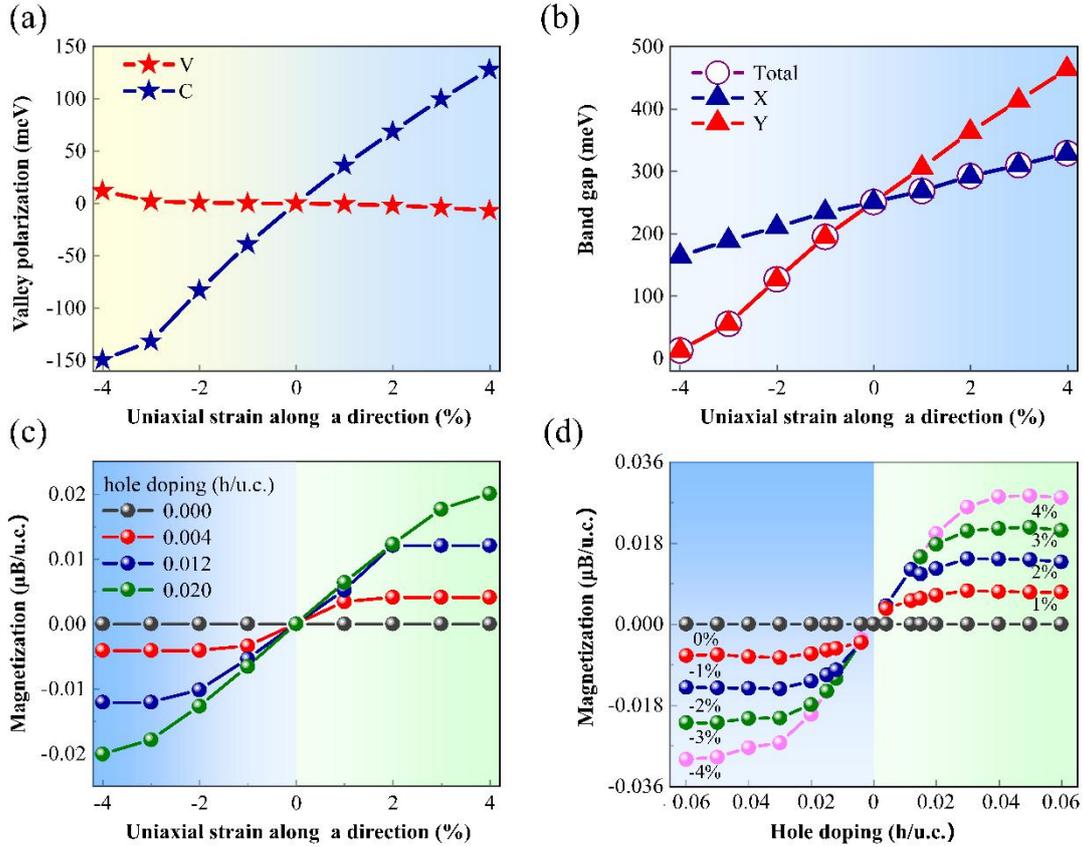

Figure 3 The strain-controlled (a) valley polarization generated at the valence band [V] and the conduction band [C], (b) valley gaps at X and Y, as well as the total bandgap, and (c)-(d) the corresponding net magnetization per unit cell for different concentration of hole doping.

Stain induced valley polarization in $V_2SeTeO$ monolayer enables new opportunities to generate net magnetic moments. Magnetism depends on integrating the spin density within the energy range from negative infinity to the Fermi level. Tuning the Fermi level to cross only one valley through carrier doping or gating can produce net moments. This is expressed as $M = \int_{-\infty}^{E_f(n)} [\rho^\uparrow(\varepsilon) - \rho^\downarrow(\varepsilon)]\, dE$, where $E_f$ is the doped Fermi level, $n$ is the doping density, and $\rho^{\uparrow(\downarrow)}$ represents the spin-up (spin-down) part of the DOS under external strain $\varepsilon$ [28].

Without doping, no net magnetic moment arises in the strained $V_2$SeTeO, consistent with the undoped band structure as shown in Figure S4. Under certain doping concentration, the magnetization increases continuously with uniaxial strain, with opposite magnetic directions induced by tensile and compressive strains. The higher doping concentrations generally produce the greater magnetization under a certain strain. The magnetization shows an initial linear response under the small strain, and eventually saturating under large strains. The higher doping concentration also leads to the higher magnetization saturation as shown in Figure 3 (c)-(d). Doping-dependent band structure of $V_2$SeTeO as a function of uniaxial strain in Figure S5, indicating the variation of magnetization polarization. Unlike noncollinear cases, piezomagnetism is rarely reported in antiferromagnets. The unique altermagnetic structure and low magnetocrystalline anisotropy of $V_2$SeTeO enable controlling the magnetic direction and moments through various means of electric field and doping.

**D. Piezoelectricity in Janus $V_2$SeTeO monolayer**

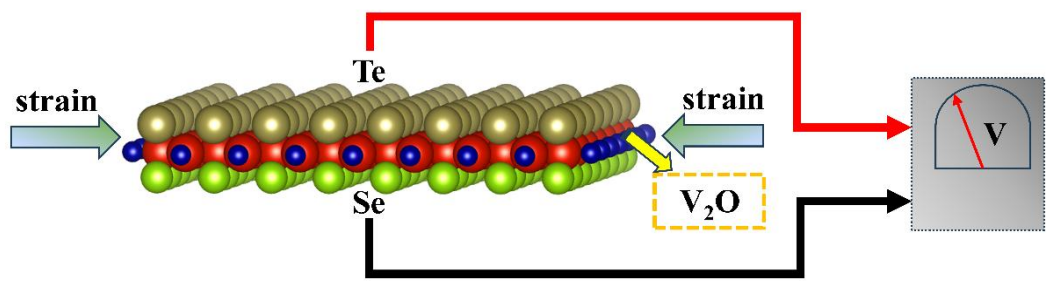

Figure 4 Schematic diagram of the mechanism for the generation of the out-of-plane piezoelectric parameter $e_{31}$, derived from the Janus structure of $V_2$SeTeO.

Piezoelectricity arises in noncentrosymmetric semiconductors, presenting electrical polarization under mechanical strain that enables electromechanical coupling. The centrosymmetric $V_2Se_2O$ monolayer lacks piezoelectricity [28], but Janus structures can introduce large out-of-plane piezoelectricity by breaking the horizontal mirror symmetry [29-32]. Substituting Se atoms in bottom layer with Te atoms breaks the horizontal mirror symmetry and induces an out-of-plane piezoelectricity in V$_2$SeTeO as shown in Figure 4. The piezoelectric tensor $e_{ijk} = \partial P_i / \partial \varepsilon_{jk}$ relates polarization $P_i$ to the strain $\varepsilon_{jk}$, where i, j, k denote the *x, y, z* Cartesian directions. This third-rank tensor characterizes the piezoelectric response [68].

Symmetry analysis indicates that the Janus V$_2$SeTeO monolayer (point group C$_{4v}$) has only one independent piezoelectric component $e_{31}$ ($e_{32} = e_{31}$). A large $e_{31}$ of 0.322*10$^{-10}$ C/m is obtained by the DFPT method, comprising opposite electronic (0.365*10$^{-10}$ C/m) and ionic (-0.043*10$^{-10}$ C/m) contributions, with the electronic part dominating. The value of $e_{31}$ in V$_2$SeTeO represents a significantly enhanced out-of-plane piezoelectricity, compared to materials like WSSe ($e_{31}$ = 0.018*10$^{-10}$ C/m) and WSeTe ($e_{31}$ = 0.010*10$^{-10}$ C/m) monolayers [32], MoSTe multilayer ($e_{31}$ = 0.119*10$^{-10}$ C/m) [32], GdClF monolayer ($e_{31}$ = -0.181*10$^{-10}$ C/m) [69], In$_2$SSe monolayer ($e_{31}$ = 0.130*10$^{-10}$ C/m) [70]. The giant vertical piezoelectric polarization is expected to enable multifunctional piezoelectric applications for Janus V$_2$SeTeO.

## IV. CONCLUSIONS

In conclusion, by means of first principles calculations we predict a novel Janus V$_2$SeTeO 2D monolayer with high dynamic and mechanical stabilities. The broken out-of-plane symmetry alongside maintained in-plane symmetry in Janus V$_2$SeTeO monolayer induces giant out-of-plane piezoelectricity while preserving original in-plane properties. This enables a remarkable multi-piezo effect combining piezovalley, piezomagnetism and piezoelectricity. Moreover, the large piezoelectric coefficient of 0.322*10$^{-10}$ C/m is found, and the valley-polarization and net magnetization of V$_2$SeTeO are larger than V$_2$Se$_2$O. Significantly, these three mechanical responses do not causally influence each other, and they are protected by different mechanisms and are modulated relatively independently, with piezoelectricity being out-of-plane, piezovalley and piezomagnetism being protected by in-plane symmetry, while the net magnetic moment is only present upon doping in the uniaxially strained V$_2$SeTeO. Our work reveals an exciting new quantum mechanical phenomenon - the multipiezo effect in a 2D Janus material. The integration of electrical, magnetic and chemical control through strain provides advantages for next-generation multifunctional nanoelectronics.

## ACKNOWLEDGEMENTS

This work was supported by National Natural Science Foundation of China (12074241, 11929401, 52120204), Science and Technology Commission of Shanghai Municipality (22XD1400900, 20501130600, 21JC1402700, 21JC1402600, 22YF1413300), China Postdoctoral Science Foundation (2022M722035), Key Research Project of Zhejiang Laboratory (2021PE0AC02), Shanghai Technical Service Center of Science and Engineering Computing, Shanghai University.

## DECLARATION OF COMPETING INTEREST

The authors declare that they have no known competing financial interests or personal relationships that could have appeared to influence the work reported in this paper.

## SUPPLEMENTARY MATERIAL

See the supplementary material for more details of the studied $V_2SeTeO$ monolayer.

# Supplementary Materials for

# Multipiezo effect in altermagnetic $V_2SeTeO$ monolayer

Yu Zhu,[1] Taikang Chen,[1] Yongchang Li,[1] Lei Qiao,[1] Xiaonan Ma,[1] Tao Hu,[2] Heng Gao,[1*] and Wei Ren[1,3‡]

[1]*Department of Physics, Shanghai Key Laboratory of High Temperature Superconductors, International Centre of Quantum and Molecular Structures, Shanghai University, Shanghai 200444, China*
[2]*School of Materials Science and Engineering, Shanghai University, Shanghai 200444, China*
[3] *Zhejiang Laboratory, Hangzhou 311100, China*

[*]gaoheng@shu.edu.cn  [‡]renwei@shu.edu.cn

[ξ]These authors contributed equally to this work.

Table S1 The structural parameters of $V_2Se_2O$ and $V_2SeTeO$ monolayers: lattice constant (Å), bond length (Å) and angle (°).

| System | Lattice Constant | Bond Length | Bond Angle |
|---|---|---|---|
| $V_2Se_2O$ | a = 4.05 Å<br>b = 4.05 Å | 4.05 Å ($V_{1/2}$-$V_{1/2}$)<br>2.66 Å (V-Se)<br>2.02 Å (V-O) | 99.07° (Se-V-Se)<br>180.00° (O-V-O)<br>90.00° (Se-V-O) |
| $V_2SeTeO$ | a = 4.08 Å<br>b = 4.08 Å | 4.08 Å ($V_{1/2}$-$V_{1/2}$)<br>2.04 Å (V-O)<br>2.65 Å / 2.85 Å (V-Se/Te) | 100.44° (Se-V-Se)<br>177.63° (O-V-O)<br>89.24°/90.83° (Se/Te-V-O) |

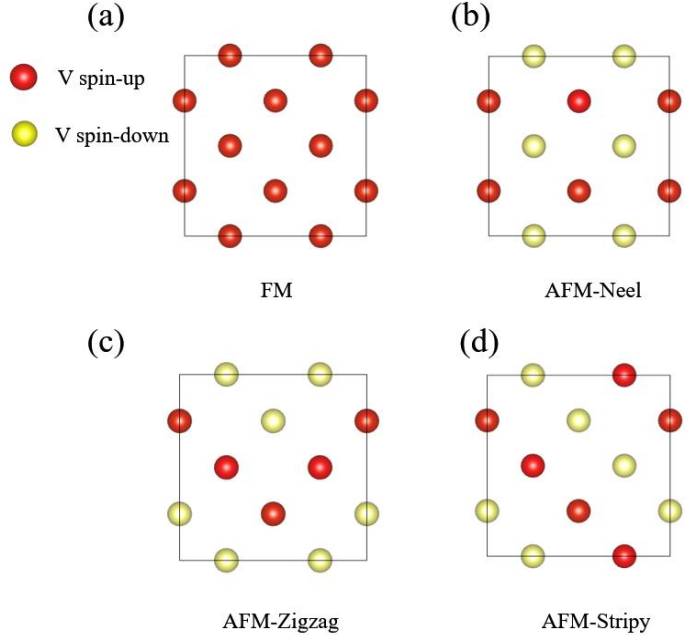

Figure S1 The initial magnetic configurations of FM and AFM for V$_2$SeTeO monolayer within a 2×2×1 supercell.

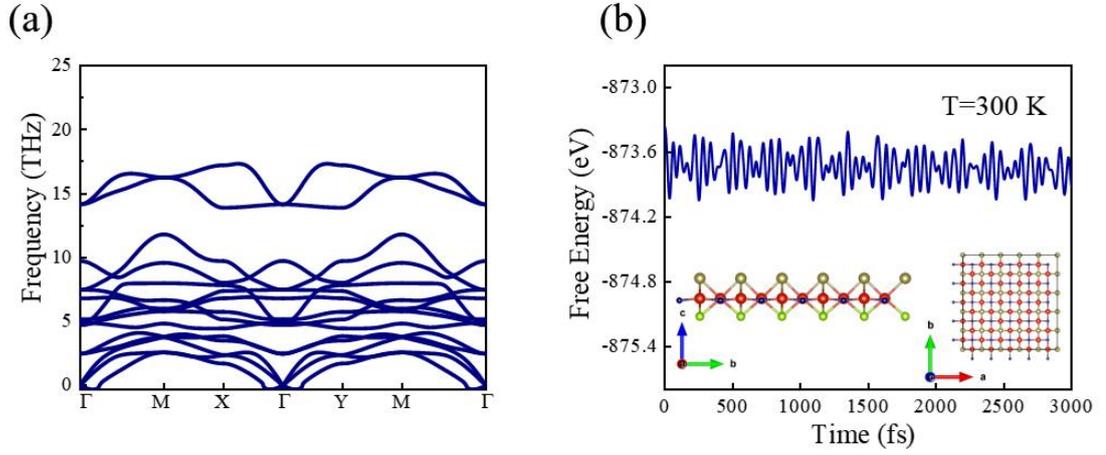

Figure S2 (a) The phonon dispersion and (b) *ab initio* molecular dynamics simulations at room temperature of V$_2$SeTeO monolayer.

Table S2 The PBE+U calculated elastic constants $C_{ij}$ (N/m), relative energies (eV per V atom) for the ferromagnetic (FM) ground state and three antiferromagnetic (AFM-Neel, AFM-Zigzag and AFM-Stripy) states for V$_2$SeTeO monolayer.

| $C_{11}$ | $C_{12}$ | $C_{66}$ | $E_{AFM-Neel}$ | $E_{FM}$ | $E_{AFM-Zigzag}$ | $E_{AFM-Stripy}$ |
|---|---|---|---|---|---|---|
| 114.18 | 18.31 | 36.87 | 0 | 0.18 | 0.13 | 0.22 |

Note S1 **The details of fundamental symmetry mechanism for altermagnetic structure:**

The PT symmetry (P means spatial inversion symmetry and T means time reversal symmetry) needs to be broken which protects spin degeneracy. However, this case is only applicable when considering spin-orbit coupling (SOC). When the SOC is turned off, the coupling effect of spin and spatial degrees is weakened, which means that there will be pure spin U, a spinor symmetry. The U symmetry reverses the spin and maintains momentum invariance, leading to preservation of spin degeneracy. Therefore, we need a further symmetry constraint, the UL symmetry breaking condition (U means spinor symmetry and L means translation symmetry). U reverses the spin state while L performs a translation operation on the primitive lattice, preserving the crystal structure and spin degeneracy.

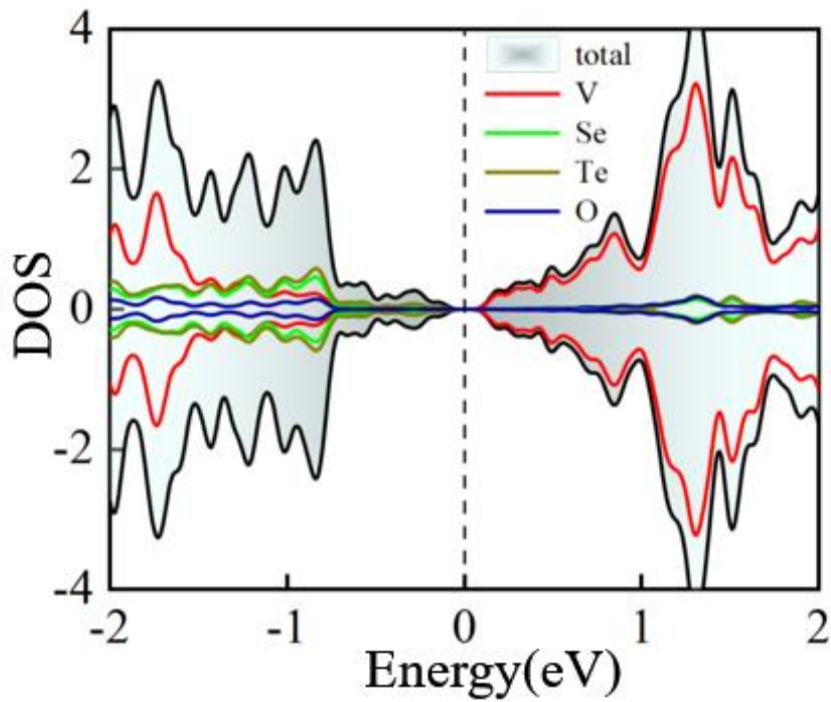

Figure S3 Partial DOS for the monolayer $V_2SeTeO$ from PBE+U calculation.

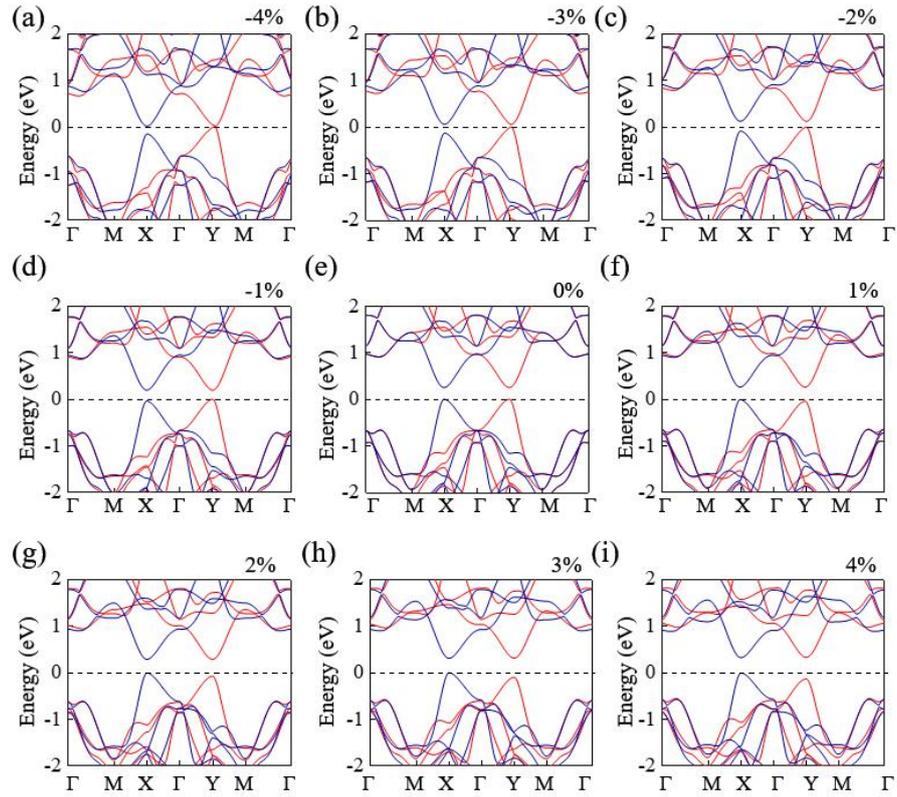

Figure S4 Spin-resolved band structures (red and blue lines for spin-up and spin-down) of the undoped V$_2$SeTeO monolayer under uniaxial strain along a direction from -4% to 4%.

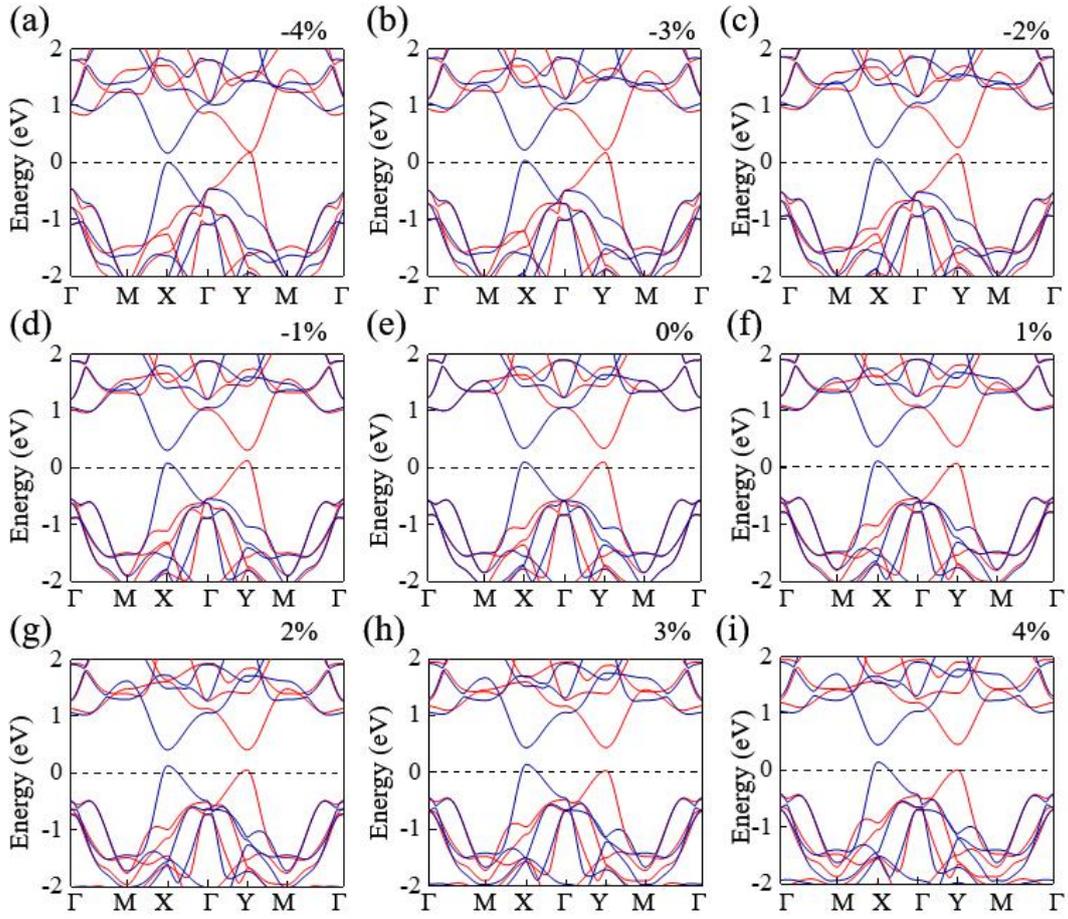

Figure S5 Spin-resolved band structures (red and blue lines for spin-up and spin-down) of monolayer $V_2SeTeO$ with the hole doping concentration of 0.02, under uniaxial strain along a direction from -4% to 4%.